\documentclass[12pt,preprint]{aastex}

\shorttitle{MHD Turbulence and Polarization of Molecular Lines}
\shortauthors{Watson and Wiebe}

\begin{document}

\title{Anisotropy of Magnetohydrodynamic Turbulence and Polarization of the
Spectral Lines of Molecules}

\author{D. S. Wiebe\altaffilmark{1} and W. D. Watson}
\affil{Physics Department, University of Illinois at Urbana-Champaign,
IL 61801}

\begin{abstract}
The anisotropy of velocities in MHD turbulence is demonstrated explicitly by
calculating the velocity gradients as a function of direction in
representative simulations of decaying turbulence.\ It follows that the
optical depths of spectral lines are anisotropic when there is MHD
turbulence, and that this anisotropy influences the polarization
characteristics of the emergent radiation. We calculate the linear
polarization that results for the microwave lines of the CO molecule in
star-forming gas and show that it is comparable to the polarization that is
observed. This and our earlier result---that the anisotropy of MHD
turbulence may be the cause for the absence of the Zeeman $\pi $-components
in the spectra of OH mainline masers---are the first demonstrations of the
occurrence of anisotropy in the optical depths caused by MHD turbulence. A
non-local approximation is developed for the radiative transfer and the
results are compared with those from a local (LVG) approximation.
\end{abstract}


\altaffiltext{1}{Permanent address: Institute of Astronomy of the RAS, 48,
Pyatnitskaya str., Moscow, 119017 Russia}

\section{Introduction}

Magnetic fields are believed to play a key role in the dynamics of the gas
of the Galaxy, including the formation of stars and related phenomena.
Dispersions in the molecular velocities are observed that exceed the thermal
dispersion, and are suggestive of turbulence. Since the gas in astronomical
environments tends to have high electrical conductivity, the turbulence is
expected to be ``MHD turbulence''--- with correlated irregularities between
the velocity fields, the magnetic fields, and the matter distributions.
Inferences about the structure of the magnetic fields in the interstellar
medium are based mostly on polarimetric observations of the attenuation of
starlight by dust and of the thermal emission by dust, and tend to reflect
the results of averaging over dimensions larger than those at which MHD
turbulence may be evident.

\cite{gk81} recognized that, if anisotropies in the optical depths do occur
for the spectral lines of molecules in the interstellar gas, they could
cause a fractional linear polarization of a few percent in the observed
spectral line radiation. More importantly, the direction of the polarization
would indicate the projected direction of the magnetic field, and would
serve to map the directions of the magnetic field lines. \cite{gk81} perform
calculations for anisotropic optical depths due to assumed anisotropies in
the molecular velocities. No specific cause for the anisotropic velocities
is given by \cite{gk81}. Anisotropic optical depths could clearly have
various origins---including, simply the proximity of the edge of the gas
cloud. The anisotropy due to a strong source of continuum radiation external
to the gas cloud could also have a similar effect in causing linear
polarization of the spectral line radiation emitted by the gas. The first
effort to detect this polarization was unsuccessful. \cite{wannier}
attempted to measure the linear polarization of thermal lines of CO, CS, and
HCN in 14 sources. Although, their upper limits were above the maximum
likely polarization calculated for the CO lines, these upper limits were
below (by factors of a few) the maximum likely polarizations calculated for
the CS and HCN lines. \cite{wannier} attributed the lack of observed
polarization to the unresolved structure of the magnetic and/or velocity
field. However, \cite{bw87} subsequently searched for polarization in an NH$%
_{3}$ line at higher resolution using the VLA---also without success.
Further, no polarization was detected for HCO$^{+}$ lines by \cite{lis88} in
four dark clouds, with an upper limit of 2\%. \cite{glenn_dr21} attempted to
measure the polarization of HCO$^{+}$ lines in DR~21, but also obtained only
an upper limit (0.4\%). Again, the absence of detectable polarization was
attributed to likely unresolved structure of the magnetic/velocity fields on
scales smaller than the telescope beamsize.

Stimulated by these non-detections, \cite{degwat84} performed multilevel
calculations that included states of higher angular momentum and
demonstrated that the polarization of a (1--0) transition is reduced by
about a factor of two. This and other considerations discussed by \cite
{lis88} reduce the expected polarization of the thermal lines to near the
upper limits of 1--2\% from the non-detections cited above.

The linear polarization of thermal lines was finally detected by \cite
{glenn97}. Polarization of a spectral line of molecular CS was detected at a
level of a few percent toward two evolved stars. \cite{glenn97} argued that
this success was a result of improved detector sensitivity and by the
relatively smooth structure of the gas in the vicinity of the evolved
star---in contrast to the clumpiness of gas in star forming regions. Linear
polarization of thermal spectral lines from the interstellar gas was
subsequently detected by \cite{greaves99}. The CO (2--1) and (3--2)
transitions in three different locations were found to have polarizations
ranging from 0.5\% to 2.4\%. The detection toward the DR~21 star forming
region was tentative, but was later confirmed by \cite{lai03} for DR~21(OH). At this time,
polarization of thermal CO lines has also been detected (or confirmed) in
the star forming regions NGC~1333 \citep{girart99}, NGC~2024 %
\citep{greaves01}, the Galactic center \citep{greaves02}, and the Orion KL
region \citep{girart04}. \cite{cortes} report the striking observation that
the polarizations of the CO (2-1) and (1-0) transitions tend to be
orthogonal in DR~21(OH), with the polarization of the (1-0) being parallel to
the polarization of the emission by the dust grains.

The polarization of the spectral lines of molecules may be useful as a
diagnostic, not only for the magnetic fields themselves, but also for the
MHD nature of the gas. One of the prominent features of MHD turbulence is
its strong anisotropy \citep{gs95}, which will enhance the magnitude of the
optical depths of spectral lines parallel to the magnetic field in
comparison with the optical depths perpendicular to the magnetic field %
\citep{ohmas}. In \cite{ohmas} we considered OH maser radiation that is
created in the presence of mildly supersonic MHD turbulence. We showed that
this anisotropy in the maser optical depths resulting from the anisotropy of
MHD turbulent velocities can suppress the Zeeman $\pi $-components of
mainline OH masers as observed---a longstanding puzzle, dating back to the
first detections of masers in astronomy.

Maser radiation probes (at least directly) only the tiny fraction of the
entire volume of a region in which maser spots are observed. In contrast,
thermal spectral lines of molecules tend to be observed over most of the
volume of molecular gas clouds. There is less doubt than for masers that the
information obtained from these spectral lines is representative of a
significant component of the gas. On the other hand, the effects of
anisotropic optical depths are much more easily recognized for the OH
mainline masers because the Zeeman $\pi $- and $\sigma $-components of these
spectral lines are well separated in frequency from one another and can
ordinarily be identified as $\sigma $'s or (if they were present) $\pi $'s.
For spectral lines of other interstellar molecules, the Zeeman splitting is
much less than the spectral linebreadth. The Zeeman components then overlap
nearly completely and are essentially indistinguishable from one another.
Nevertheless, an anisotropy in the optical depths is the origin of the Goldreich-Kylafis
effect and, though its effect is small, can be detected. The goal of this
Paper is to assess whether the anisotropy of plausible MHD turbulence is
sufficient to cause the observed linear polarization of the spectral lines
of CO from interstellar gas clouds that is observed. In our calculation for
the OH masers, it was only necessary to integrate through a gas in which the
molecular populations are independently specified in order to obtain the
intensities of the maser rays. Calculating the polarization of the thermal
lines of CO is more challenging. The populations of the magnetic substates
must first be found by solving rate equations with the anisotropic fluxes
derived from the anisotropic optical depths.

In \S ~2, we calculate the velocity gradients within representative examples
of turbulent MHD gas that are obtained from numerical simulations of
decaying turbulence and demonstrate the anisotropy as a function of the
relevant parameter here---the ratio of the Alfv\'{e}n velocity to the sound
speed of the gas. The existing calculational methods for the
Goldreich-Kylafis effect must be augmented somewhat because of the variation
in the properties of the gas from point to point, which we wish to consider.
We describe how the calculations are performed in \S ~3. Calculations are
performed, not only for the average polarization, but also to indicate how
the polarization can change in direction and magnitude across the face of a
gas cloud and across the profile of the spectral line. The results of these
calculations are presented in \S ~4. A summarizing discussion is provided in
\S ~5 where we relate the results here to our previous calculations for the
polarization of continuum emission from dust grains aligned by irregular
magnetic fields.

\section{The Anisotropy of Velocities in an MHD Gas}

Our quantitative analysis of the anisotropy is based on a numerical model of
decaying MHD turbulence. We use the same results of simulations as were used
in \cite{ohmas}. Only a brief description of these simulations is provided
here, and we refer the reader to the earlier paper and to references therein
for details. Velocity fields and magnetic fields were obtained there by
integrating the equations of compressible, ideal MHD turbulence in a cubic,
periodic domain on a uniform grid with $128^{3}$ cells. The key parameter
that characterizes these simulations is the ratio of the Alfv\'{e}n velocity
to the speed of sound $v_{\mathrm{A}}/c_{\mathrm{s}}$. This Alfv\'{e}n
velocity is calculated from the average magnetic field in the MHD
cube---which remains constant during the time evolution of the simulation.
Three values of $v_{\mathrm{A}}/c_{\mathrm{s}}$ are considered---1, 3 and
10. As is standard for such simulations, the computations begin with
plausible, initial velocity perturbations and evolve with time. To verify
that our conclusions do not depend upon a particular statistical choice for
these initial perturbations, computations were performed for three
independent choices for these initial velocity perturbations for each of the
values of $v_{\mathrm{A}}/c_{\mathrm{s}}$. In these simulations, the
turbulent velocities evolve from supersonic (Mach number $M\simeq 4$) to
subsonic ($M\simeq 0.3$). For each combination of initial perturbations and
value for $v_{\mathrm{A}}/c_{\mathrm{s}}$, we have analyzed the fields at
nine time steps separated in time by intervals of $0.2L_{0}/c_{s}$---where $%
L_{0}$ is the physical length to be associated with the 128 grid points
along the edge of the computational cube.
\clearpage
\begin{figure}[tbp]
\includegraphics[width=0.8\textwidth]{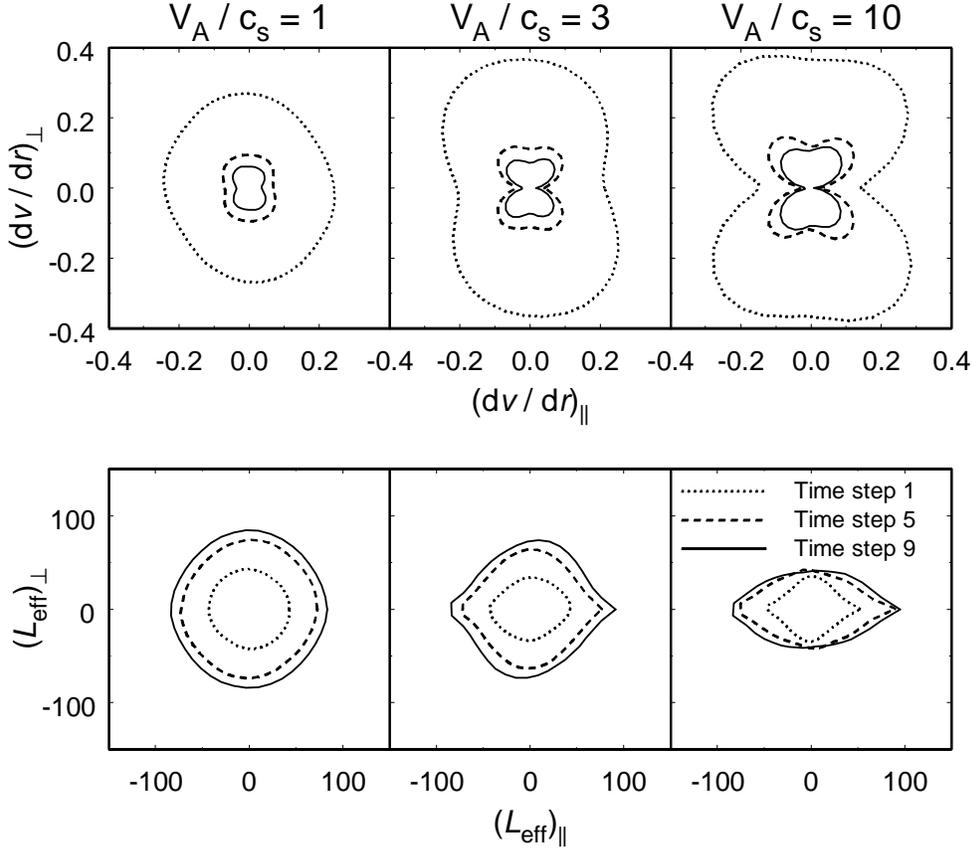}
\caption{Top panel---the angular distribution of velocity gradients for
three $v_{\mathrm{A}}/c_{\mathrm{s}}$ values at three time steps. The
distance from the coordinate origin to the curve is equal to the
cube-averaged velocity gradient in the corresponding direction. Velocity is
measured in units of the thermal velocity of the gas $v_{\mathrm{th}}$, and
the length is measured in units of the number of grid cells. Bottom
panel---the angular distribution of effective interaction lengths (see text)
that incorporate non-local radiative interactions between the populations of
the gas molecules. The length is measured in units of the number of grid
cells. The subscripts $||$ or $\perp $ indicate directions parallel or
perpendicular to the average magnetic field.}
\label{butterfly}
\end{figure}
\clearpage
Anisotropy in the optical depths of spectral lines in an MHD gas has already
been demonstrated in \cite{ohmas}. Here we delineate the anisotropy of the
medium more explicitly and in more detail. There are different, approximate
ways to characterize this anisotropy. We might assume that, in a
supersonically turbulent medium, only the nearby region influences the
populations at a given location so that the large velocity gradient (LVG)
approximation is appropriate. At each of the grid points of the MHD cubes,
the anisotropy can then be visualized in terms of the angular distribution
of the radial (relative to the location of the particular grid point)
gradients of the radial component of the turbulent velocity 
\begin{equation}
\mathrm{d}v_{r}/\mathrm{d}r.
\end{equation}
To examine the anisotropy in this approximation, we compute these gradients
(using linear interpolations of velocities at neighboring locations) at a large 
number of grid points and in a large number of directions at each grid point. To simplify the computations, we only consider directions at each grid point that lie within a single plane. A plane is specified as containing the direction of the average magnetic field and the particular grid point. The planes associated with the various grid points are all parallel. The results for all grid points are then combined  to obtain 
averaged gradients that are a function of the angle of inclination relative to the 
direction of the average magnetic field in the cube.  These gradients are presented
in Figure~\ref
{butterfly} (top panel) for the three representative time steps and for the
three values of $v_{\mathrm{A}}/c_{\mathrm{s}}$ that we consider. The
gradients in Figure~\ref{butterfly} are based on only a single sequence of
time steps (resulting from one of the three ensembles of initial
perturbations) for each of the values of $v_{\mathrm{A}}/c_{\mathrm{s}}$.
For the same $v_{\mathrm{A}}/c_{\mathrm{s}}$ and the same time step, the
averaged gradients for the other MHD cubes differ negligibly from those that
are presented.

In a statistical sense, the properties of such MHD cubes should be symmetric
about the direction of the average magnetic field. The curves in Figure~\ref{butterfly}
deviate slightly from this exact symmetry because they are based on only
a single MHD cube, and hence, on only a single ensemble of initial 
perturbations. 

The direction of the average magnetic field is parallel to the horizontal
axis in Figure~\ref{butterfly}. As expected, the velocity gradients are
smallest in directions parallel to the magnetic field and largest in
directions that are at large angles to the magnetic field. The angular
variations of the gradients in Figure~\ref{butterfly} can be seen to be
similar at all times in the simulation for a specific value of $v_{\mathrm{A}%
}/c_{\mathrm{s}}$. In decaying turbulence, the turbulent velocities decrease
with time, and hence the magnitudes of these gradients also decrease.
Anisotropy is clearly evident in Figure~\ref{butterfly} for $v_{\mathrm{A}%
}/c_{\mathrm{s}}=3$ and 10, though not for $v_{\mathrm{A}}/c_{\mathrm{s}}=1$%
. Note that the magnitudes of the velocity gradients in Figure~\ref
{butterfly} are given in terms of the thermal velocity of the gas $v_{%
\mathrm{th}}$ ($=\sqrt{2kT/m}$)---the thermal velocity of the dominant
atomic or molecular component of the gas. The gas is mainly H$_{2}$ in the
astronomical environments where our calculations are expected to be applied
and the molecules such as CO from which the spectral line radiation of
interest arises are much heavier than H$_{2}$. Hence, the numerical values
of the velocity gradients in Figure~\ref{butterfly} (top panel) should be
divided, for example, by the ratio (CO mass/H$_{2}$ mass)$^{1/2}=3.74$ to
obtain the quantities that are relevant for the radiative transport of the
spectral lines of the CO molecule in the LVG approximation (see, e.g.,
equation \ref{lomega}). Note also that the dimensionless distance scale is
the separation between grid points so that the gradients in Figure~\ref
{butterfly} should be divided by ($L_{0}/128$) to express them in terms of
ordinary units for distance.

In a turbulent medium, we may expect that some remote points (as well as the
nearby points) contribute to the average intensity of radiation at a given
location and thus, influence the populations at this location. In contrast,
in the LVG approximation where only `local' interactions are considered, the
extent of the region that influences the populations at a given location is
assumed to be proportional to the inverse of the velocity gradient. That is,
in the LVG approximation the `interaction length' is assumed to be the
distance over which the velocity would change by $v_{\mathrm{th}}$ in a
given direction if its gradient were constant along this direction. In order
to characterize the non-local, radiative coupling of the molecular
populations and to see whether the anisotropy is preserved in the
`non-local' case as well, we introduce `effective' interaction lengths 
\begin{equation}
L_{\mathrm{eff}}=\int\limits_{0}^{L_{\max }}\mathrm{d}r\exp \left[ -14\left(
v_{c}-v_{r}\right) ^{2}\right] ,  \label{leff}
\end{equation}
which are computed in particular directions from the specified location.
Here $v_{c}$ is the component of the local turbulent velocity in the
direction of the integration path, $v_{r}$ is this component of the
turbulent velocity at a distance $r$ from the relevant point, and the factor
14 is the ratio of molecular masses CO/H$_{2}$. In the local case, the
interaction length is defined by the velocity gradient and all locations
along the length of the ray have equal weight. In the non-local case, we
define a length $L_{\max }$ of a ray along which the optical depths are to
be calculated. Locations along this ray have different weights as determined
by the Gaussian profile function in the integrand of equation (\ref{leff}).

The cube-averaged, angular distributions of $L_{\mathrm{eff}}$ are also
shown in Figure~\ref{butterfly} (bottom panel) for the choice $L_{\max }=100$
grid spacings. Again, there is almost no anisotropy in the $v_{\mathrm{A}%
}/c_{\mathrm{s}}=1$ case, whereas the anisotropy is evident in the plots for
the $v_{\mathrm{A}}/c_{\mathrm{s}}=3$ and 10 cases. Since the interaction
length is similar to the inverse of the velocity gradient, it is the largest
in the direction of the magnetic field. The size of the interaction region
increases with time as the velocity dispersion in the cube decreases. The
degree of anisotropy also increases somewhat. Despite the smooth appearance
of the averaged curves, there are significant irregularities at individual
locations as we will show in \S ~4. Just 
as for the average velocity gradients, the interaction lengths in Figure~\ref{butterfly} are not exactly symmetric 
about the direction of the average magnetic field because Figure~\ref{butterfly} is based on the evolution 
of an individual MHD cube.

\section{Methods for Calculating the Polarized Radiation}

The goal here is to calculate the linear polarization of spectral line
radiation that emerges from a gas cloud with irregular magnetic and velocity
fields represented by the MHD fields from the simulations described above.
The focus is on incorporating the variations in the fields from location to
location within the gas. That is, instead of using a single, averaged
angular distribution for the entire cube (such as those in Figure \ref
{butterfly}) to represent the anisotropy caused by the magnetic fields and
velocities in the gas, we calculate the relevant angular distribution at
each grid point using the actual magnetic fields and velocities from the
simulations at the particular grid point. In addition to being more
realistic, this will allow us to assess what differences in the polarization
characteristics can be expected for radiation that emerges from different
locations on the surface of a gas cloud. The calculations are performed for
the specific case of the $J=1\rightarrow 0$ rotational transition of the CO
molecule with the usual assumption that the results will be indicative for
other transitions.

The radiative transfer equations for the Stokes parameters are integrated
through the cube for specific rays of spectral line radiation that emerge
from the MHD cube and represent the observed radiation. For a $%
J=1\rightarrow 0$ transition with a resonant frequency $\nu _{0}$, these can
be expressed as 
\begin{equation}
{\frac{\mathrm{d}}{\mathrm{d}s}}\left( 
\begin{array}{c}
I_{v} \\ 
Q_{v} \\ 
U_{v}
\end{array}
\right) =\left[ -\left( 
\begin{array}{ccc}
A & B & 0 \\ 
B & A & 0 \\ 
0 & 0 & A
\end{array}
\right) \left( 
\begin{array}{c}
I_{v} \\ 
Q_{v} \\ 
U_{v}
\end{array}
\right) +\left( 
\begin{array}{c}
k_{\pm }S_{\pm }+k_{0}S_{0} \\ 
k_{\pm }S_{\pm }-k_{0}S_{0} \\ 
0
\end{array}
\right) \right] \phi (\nu -\nu _{0}\frac{v_{s}}{c}),  \label{transfer}
\end{equation}
where 
\begin{equation}
\begin{array}{ll}
A=(k_{\pm }+k_{0})/2, &  \\ 
B=(k_{\pm }-k_{0})/2, & 
\end{array}
\end{equation}
$\phi $ is the normalized, Gaussian line profile caused by the Doppler
shifts associated with a thermal distribution of molecular velocities, $v_{s}$ is the line-of-sight component of the
turbulent velocity at the location $s$ along the ray, and the absorption coefficients and source
functions ($k_{\pm },k_{0},S_{\pm },S_{0}$) are obtained from the populations of the magnetic
substates. Explicit expressions for ($k_{\pm },k_{0},S_{\pm },S_{0}$) can be
found e.g., in \cite{degwat84} [but see \cite{DWW} for some typographical
corrections]. The populations $n$ of the magnetic substates at a particular
location along a ray are found by solving the usual rate equations in steady
state 
\begin{equation}
0={\frac{\mathrm{d}n_{i}}{\mathrm{d}t}}=-A_{\mathrm{E}%
}n_{i}+R_{i}(n_{j}-n_{i})+(C_{ji}n_{j}-C_{ij}n_{i}),  \label{pops}
\end{equation}
where $A_{\mathrm{E}}$ is the Einstein ``A-coefficient'' for spontaneous
emission, and $i$ ($=\pm 1,0$) and $j$ ($=0$) refer to the magnetic
substates of the upper $J=1$ and lower $J=0$ energy levels, respectively. In
writing the above equations, the quantization axis and the axis for defining
the polarizations are both assumed to be along the direction of the magnetic
field. This simplification is appropriate when the Zeeman splitting is much
less than the inverse lifetime of the excited molecular state---a
requirement that is well satisfied for CO and most molecules in the
interstellar gas. While the average magnetic field is always parallel to one
of the sides of the cube, the local magnetic field fluctuates around this
average direction. Thus, in solving equations (\ref{transfer}) and (\ref
{pops}), it is necessary to transform Stokes $Q_{\nu }$ and $U_{\nu }$ by a
coordinate rotation at each grid point so that they are defined relative to
the direction of the local magnetic field. Note also that the circular
polarization is assumed to be negligible. Circular polarization is
negligible when the Zeeman splitting is much less than the spectral
linebreadth and the fractional linear polarization is small.

We will find the intensities to be used in calculating the $R_{i}$ by two
separate methods. Both are approximate and, especially for the second
method, are in the spirit of approximations introduced recently by others %
\citep{ossenkopf}. The populations must be computed by iteration with the
above equation at a large number of grid points. To find the $R_{i}$ which
enter into equation (\ref{pops}), the intensities at each grid point must be
determined in a large number of directions to perform the integration over
angles. So that the calculations are manageable for us, we make the
approximation that the populations that enter into the calculation of the
intensities for the $R_{i}$ are the same as at the specific grid point for
which the $R_{i}$ are being calculated. Then, Stokes $U$ is zero for the
intensities that enter into the $R_{i}$, and the $R_{i}$ can be expressed as %
\citep{degwat84} 
\begin{equation}
R_{0}=\frac{3A_{\mathrm{E}}c^{2}}{2h\nu _{0}^{3}}\int \frac{\mathrm{d}\Omega 
}{4\pi }\int \mathrm{d}\nu \,\phi \left( \nu -\nu _0\frac{v_{s}}{c}\right)
\sin ^{2}\gamma I_{\nu }^{||}(\Omega )  \label{R0}
\end{equation}
and 
\begin{equation}
R_{\pm }=\frac{3A_{\mathrm{E}}c^{2}}{4h\nu _{0}^{3}}\int \frac{\mathrm{d}%
\Omega }{4\pi }\int \mathrm{d}\nu \,\phi \left( \nu -\nu _0\frac{v_{s}}{c}%
\right) \left[ I_{\nu }^{\perp }(\Omega )+\cos ^{2}\gamma I_{\nu
}^{||}(\Omega )\right] .  \label{R1}
\end{equation}
Here the superscripts $||$ or $\perp $ designate linear polarizations
parallel or perpendicular to the magnetic field.

The first approximate method for finding the intensities to evaluate $R_{0}$
and $R_{\pm }$ is the conventional LVG approximation. To apply the
approximation it must be assumed that the turbulence creates large,
macroscopic velocity variations in most parts of a cube, and hence that
remote points of the gas do not interact significantly with each other
through the emission and absorption of spectral line radiation. The volume
that is radiatively coupled to a specific point is defined by an interaction
length. The interaction length $L(\Omega )$ in the direction specified by
the solid angle $\Omega $ is defined as the inverse of the velocity gradient
in this direction, with the velocity expressed in units of the thermal
velocity $v_{\mathrm{th}}$
\begin{equation}
L(\Omega )={\frac{v_{\mathrm{th}}}{\mathrm{d}v_{r}/\mathrm{d}r}}.
\label{lomega}
\end{equation}
The specific intensities $I_{\nu }^{\perp }(\Omega )$ and $I_{\nu
}^{||}(\Omega )$ at each grid point are then computed in the LVG
approximation using the escape probability $\beta $ evaluated for the
optical depths 
\begin{equation}
\tau ^{q}(\Omega )=k^{q}L(\Omega )
\end{equation}
at each grid point, where $q$ is either $||$ or $\perp $. The populations
entering the expressions for $k^{q}$, are assumed to be those at the
specific point for which the $R_{i}$ are being calculated. The values of $%
L(\Omega )$ at each grid point and in each direction $\Omega $ are computed
from the actual velocity field in the cube of MHD velocities (typically,
1600 directions are used at each grid point for the angular integration).
Iterations are performed to obtain consistent values for the populations and
for the radiative rates. Explicit expressions for the polarized intensities
in the LVG approximation can be found in, e.g., \cite{degwat84}.

In the second approximate method---which we designate as the non-local
approximation, we take into account possible radiative coupling between
distant points. We integrate the formal solution of the radiative transfer
equations 
\begin{equation}
I_{\nu }^{q}(\Omega )=\int\limits_{0}^{L}\,k_{l}^{q}S_{l}^{q}\phi \left( \nu
-\nu _{0}\frac{v_{l}}{c}\right) \exp \left[ -\tau _{\nu }(l)\right] \mathrm{d%
}l+I_{\nu }^{\mathrm{bg}}(\Omega )\exp \left[ -\tau _{\nu }(L)\right] .
\label{formal}
\end{equation}
where the optical depth is 
\begin{equation}
\tau _{\nu }^{q}(l)=\int\limits_{0}^{l}\mathrm{d}l^{\prime }\,k_{l^{\prime
}}^{q}\,\phi \left( \nu -\nu _{0}\frac{v_{l^{\prime }}}{c}\right)
\label{optdepth}
\end{equation}
by again assuming that the relative populations along the path of the
integration are the same as at the particular grid point where the $R_{i}$
are being computed. These relative populations are scaled with the varying
gas density in the MHD cubes before performing the integrations. Here $%
I_{\nu }^{\mathrm{bg}}(\Omega )$ is the intensity of the background
radiation that may include a contribution from an external source. The
absorption coefficient and the source function $k_{l}^{q}$ and $S_{l}^{q}$
are evaluated at the position $l$ along the particular ray. The length $L$
is taken to be approximately the cube dimension $L_{0}$ regardless of the
location of the specific grid point within the MHD cube. Since the MHD
simulations adopt periodic boundary conditions, we use the periodicity to
extend the MHD cube when necessary so that all integrals for the $I_{\nu
}^{q}$ can be computed for $L\approx L_{0}$. Just as in the first (the LVG)
approximate method, the rate equations are then solved iteratively to find
the populations $n_{\pm }$ and $n_{0}$. Note that equation (\ref{formal})
must be solved in detail at each point in this non-local approximation for a
number of frequencies $\nu $ (typically, 31 and up to 91 in some test
cases), as well as angles $\Omega $ (typically, 600 at each grid point), to
perform the integrations in equations (\ref{R0}) and (\ref{R1}). Hence, the
non-local method requires considerably more effort than does the LVG
approximation, for which a detailed calculation at a number of frequencies
is not required.

The populations at the grid points of the MHD cube obtained with either the
first or the second method described above are then used in the radiative
transfer equation (\ref{transfer}), which is integrated to obtain the
intensities for a number of parallel rays that pass through the MHD cube and
emerge at the locations of grid points on the surface. These rays represent
the emergent radiation from a gas cloud.

\section{Results}

Calculations for the emergent, polarized radiation are performed for all of
the cubes of MHD velocities and magnetic fields that are available from the
simulations---cubes at nine time steps in the evolution with the three
statistically chosen sets of initial velocity perturbations for each of the
three values of $v_{\mathrm{A}}/c_{s}$ for a total of 81 MHD cubes. In all
of the results that are presented, the line of sight is perpendicular to the
average magnetic field. We have verified that, as expected, no significant
polarization is produced in our calculations when the line of sight is
parallel to the average magnetic field (the polarization is not exactly zero
because of the irregularities in the magnetic field).

Once the specific MHD cube is selected, the calculations depend on only two
additional quantities---the gas density and the average column density of CO
molecules through the cube $N_{\mathrm{CO}}$. The latter is, specifically, $%
N_{\mathrm{CO}}=n_{\mathrm{CO}}\times L_{0}$ where $n_{\mathrm{CO}}$ is the
average density of CO molecules. The coefficients $C_{ij}$ in equation (\ref
{pops}) reflect the influence of collisional excitations and de-excitations,
and are proportional to the gas density. The gas is mainly H$_{2}$, so we
use rate coefficients for CO-H$_{2}$ collisions. The calculation actually
depends on only the ratios $C_{ij}/A_{\mathrm{E}}$. Hence, we use this
ratio---which we write as $C/A$---instead of the gas density to specify the
parameters on which the calculations depend. The value of $C_{ij}$ for
de-excitation is used to specify $C/A$. For reference purposes, $C/A=0.1$
for an H$_{2}$ density of 228 cm$^{-3}$ at a temperature of 30\thinspace K.
The polarization characteristics are insensitive to the adopted gas
temperature, as well as to the exact values of the rate coefficients for
collisional excitation. As mentioned above, the length $L$ of a ray for the
non-local approximation in equation (\ref{formal}) is taken to be equal to
the cube size. We tried other values for $L$ as well, ranging from
approximately half the cube size to $2L_{0}$, and found that the fractional
polarizations for specific rays are changed by no more than 30\%. The
highest fractional polarizations are produced at $L\approx L_{0}$.
\clearpage
\begin{figure}[tbp]
\includegraphics[scale=.70]{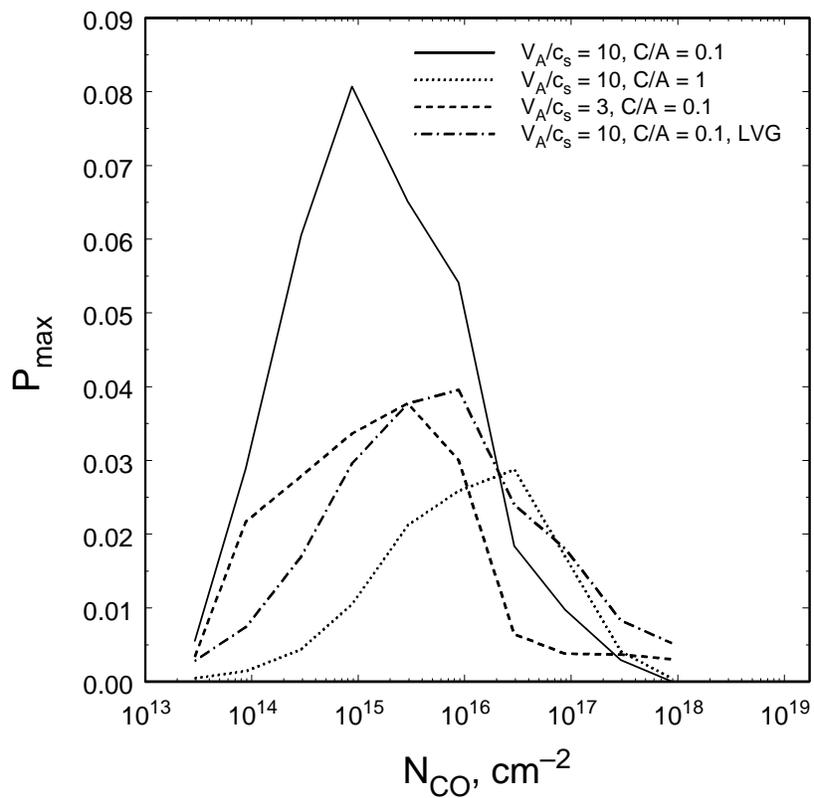}
\caption{Fractional polarization $P_{\mathrm{max}}$ at the peak intensity of
the CO spectral line as a function of the CO column density for a few
choices of $v_{\mathrm{A}}/c_{s}$ and the gas density as indcated by $C/A$ (5th time step). The solid,
dotted, and dashed lines are obtained with the non-local approximation. The
dash-dotted line is obtained with the LVG approximation.}
\label{pvstau}
\end{figure}
\clearpage
The overall dependence of the magnitude of the fractional polarization on
the CO column density is indicated in Figure~\ref{pvstau}. There, the
fractional polarization at the peak intensity for the brightest ray of the
(128)$^{2}$ rays that emerge from the grid points on the face of the cube is
shown. The maximum degree of polarization occurs at a column density of $%
\sim 10^{15}$~cm$^{-2}$ which at this density and temperature corresponds
roughly to $\tau \sim 1$---the value usually assumed to be most favorable
for the Godreich-Kylafis effect. Optical depth enters into the expressions
for the radiative rates $R_{i}$ in the rate equations, as well as in the
integration of equation (\ref{transfer}) along the line of sight for the
emergent intensities. The curves in Figure~\ref{pvstau} are reminiscent of
similar diagrams for the pure LVG computations 
\citep[e.g.][their
Fig.~1]{degwat84}. The fractional polarization is higher for the lower $C/A$%
, reaching 3\% at $C/A=1$ and 8\% at $C/A=0.1$ for the strong magnetic field
case. When the average magnetic field is lower, the fractional polarization
also is smaller (by about a factor of 2 here) as expected, indicating less
anisotropy in the medium.

\begin{figure}[ptb]
\includegraphics{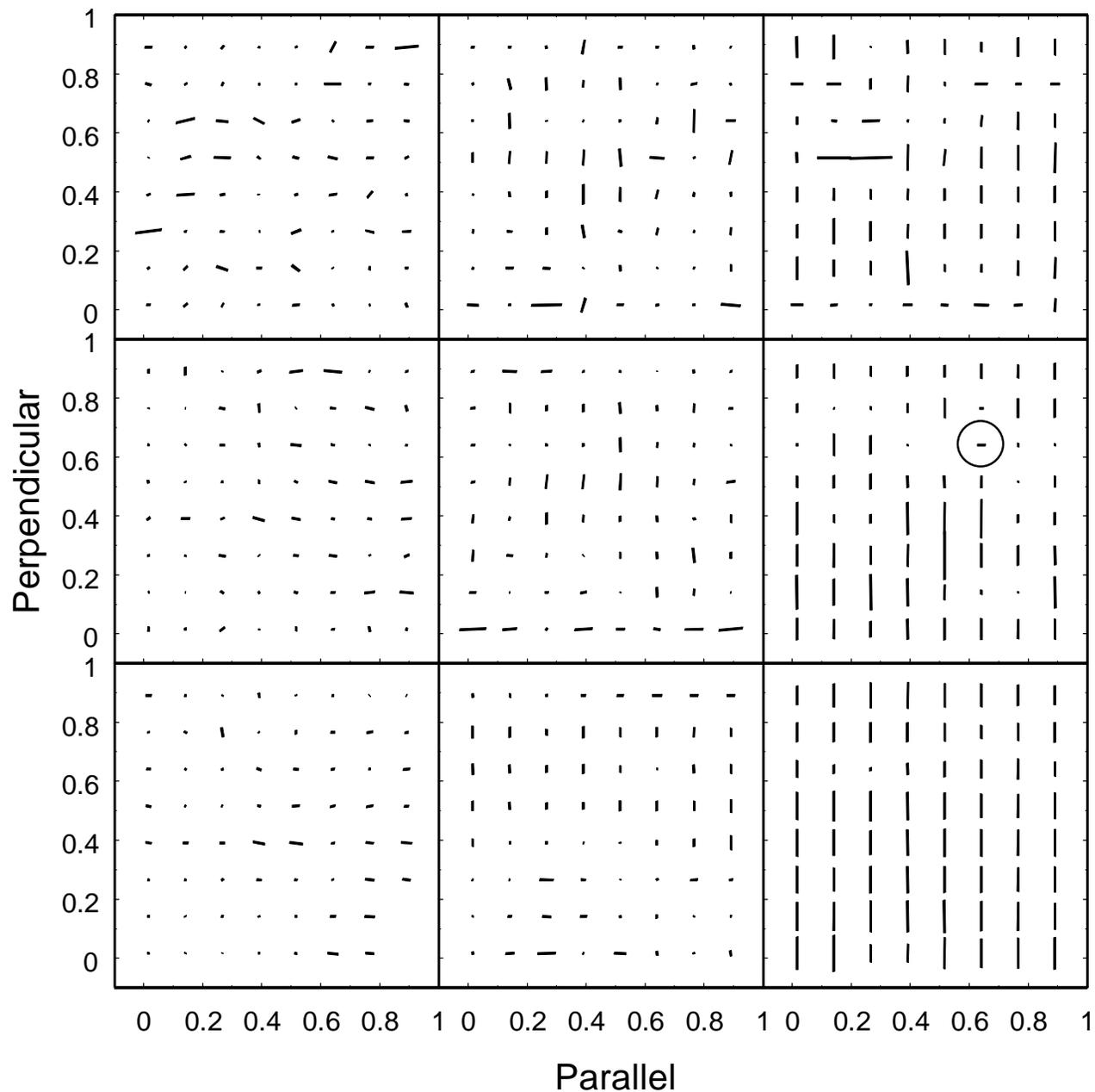}
\caption{Representative maps of polarization vectors at the maximum
intensity in the CO spectral line for various $v_{\mathrm{A}}/c_{s}$ values
(1, 3, and 10 from left to right) and time steps (3rd, 5th, and 9th from top
to bottom, in units of $0.2L/c_{s}$) obtained with the non-local
approximation. The labels for the axes refer to directions parallel and
perpendicular to the average magnetic field. The longest vector corresponds
to the fractional polarization of 5.4\%. Each 16th ray in both directions is
shown. A circle marks the ray, for which line profiles are shown in Figure~%
\ref{prof015}.}
\label{vecmapnl}
\end{figure}

\begin{figure}[tbp]
\includegraphics{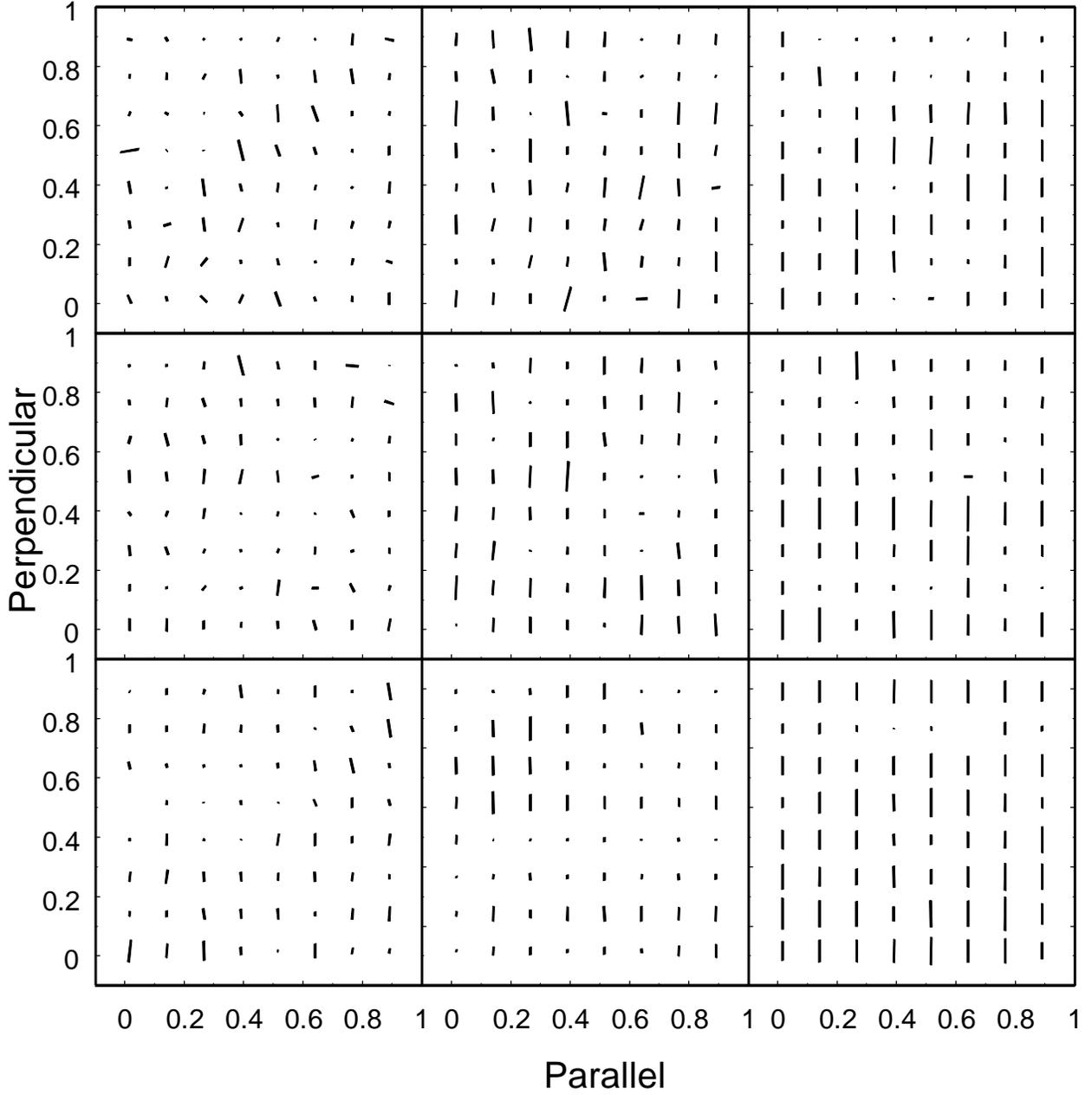}
\caption{ Same as in Figure \ref{vecmapnl}, but obtained with the LVG
approximation. The longest vector corresponds to the fractional polarization
of 4.0\%.}
\label{vecmaplvg}
\end{figure}

Another manifestation of the increase in anisotropy with stronger average
magnetic fields is provided by a comparison of the representative maps of
the fractional polarization at the three values $v_{\mathrm{A}}/c_{\mathrm{s}%
}=1$, 3, and 10 for three representative time steps in the evolution of the
turbulence (Figure~\ref{vecmapnl} and \ref{vecmaplvg}). The linear dimension
of the cube is taken to be 0.12~pc, which corresponds to $N_{\mathrm{CO}%
}\approx 10^{16}$~cm$^{-2}$ for the adopted values of $C/A\approx 0.1$ and
the assumed CO abundance ($10^{-4}$ relative to H$_{2}$). We use this column
density, which is an order of magnitude higher than the density that
provides the highest fractional polarization (see Figure~\ref{pvstau}),
because it produces brighter lines. These lines will be more easily
observable, and will still have appreciable fractional polarization.

The Figures differ in that the non-local approximation is used in Figure~\ref
{vecmapnl} to calculate the molecular populations at each grid point whereas
the LVG approximation is used to calculate the populations in Figure~\ref
{vecmaplvg}. In the regime here where the Zeeman shift is much greater than
the collision rate $C_{ij}$ or the radiative decay rate $A_{\mathrm{E}}$,
the direction of the spectral line polarization that is generated at a grid
point will tend to be either parallel or perpendicular to the magnetic
field\ at that grid point. Whether the polarization is parallel or
perpendicular is, in the absence of a strong external source of radiation,
determined by the angle between the local velocity gradients and the
magnetic field. Thus, the appearance of the polarization maps is defined by
both the magnetic field structure and the velocity field structure. Since we
know from Figure~\ref{butterfly} that the average velocity gradients are
strongest perpendicular to the magnetic field, the preferred direction for
the polarization vectors should be perpendicular to the magnetic field %
\citep[e.g.,][]{gk81}. Relative to the direction of the average magnetic
field, this is exactly what is seen in our computations. Because of the
turbulence, the actual magnetic fields at many grid points will be in
directions that are different from the average. We might then expect that
the direction of the polarization of the radiation that is generated at
these grid points will be neither parallel nor perpendicular to the
direction of the average magnetic field. As a result, polarization
directions might be expected in Figures~\ref{vecmapnl} and~\ref{vecmaplvg}
that are neither exactly parallel or perpendicular to the average magnetic
field in the cube, which is along the horizontal axes in the Figures. That
such polarization directions are not more common in these Figures is then
noteworthy---especially for $v_{\mathrm{A}}/c_{\mathrm{s}}=1$ where the
variation in the directions of the local magnetic fields are greatest since
the turbulent component of the magnetic field is strongest in comparison
with the average magnetic field.
\clearpage

\clearpage
A few trends are evident in Figures~\ref{vecmapnl} and~\ref{vecmaplvg}. The
fractional polarizations are greater in models with stronger magnetic
fields, where we expect the anisotropy related to the MHD turbulence to be
greatest. Also, the pattern of the polarization directions is more regular
in the models with the stronger magnetic fields. In the $v_{\mathrm{A}%
}/c_{s}=1$ case, the polarization vectors do reflect the presence of the
significant, irregular component of the magnetic field. In models with the
stronger average magnetic fields (i.e., larger $v_{\mathrm{A}}/c_{s}$), the
only variations in the directions of the polarization that persist are at
locations where the polarization directions are parallel to the average
magnetic field, instead of perpendicular to it. The number of these
polarization ``reversals''---where the polarization direction is parallel
rather than perpendicular to the average magnetic field---constitutes almost
one third of the total number of rays that are considered for the non-local
computations when $v_{\mathrm{A}}/c_{s}=3$ , while only a few reversals are
seen the $v_{\mathrm{A}}/c_{s}=10$ case. Also, in the latter case the number
of reversals decreases with time. Reversals indicate that, despite the
general tendency for the velocity gradients to be perpendicular to the
average direction of the magnetic field, there are some locations where they
are mostly parallel to the average magnetic field.

We have performed computations in which the irregularities in the magnetic
fields (but not in the velocities) are artificially removed, and obtained
essentially the same results. We thus conclude that the reversals are mainly
due to the changes in direction of the velocity gradients.

Figures~\ref{vecmapnl} and~\ref{vecmaplvg} do not agree well in detail on a
point-by-point basis. The LVG maps are more regular, and have a smaller
number of reversals. The profiles for the emergent intensities that are
calculated for the LVG approximation and for the non-local approximation are
somewhat different when the line is optically thin, even though they almost
coincide when the column density exceeds $10^{16}$~cm$^{-2}$ ($\tau \ga10$).
While the Stokes-Q profiles for the two approximations are also similar at
most locations, there are some points where the specific velocity structure
causes the polarization vector to be parallel to the average magnetic field
in the non-local case and perpendicular to it in the LVG case or vice versa.
We nevertheless interpret the general agreement in the character and in the
average magnitude of the polarization obtained with the two approximations
as support for our working hypothesis that these approximate methods provide
useful results.
\clearpage
\begin{figure}[tbp]
\includegraphics[scale=.80]{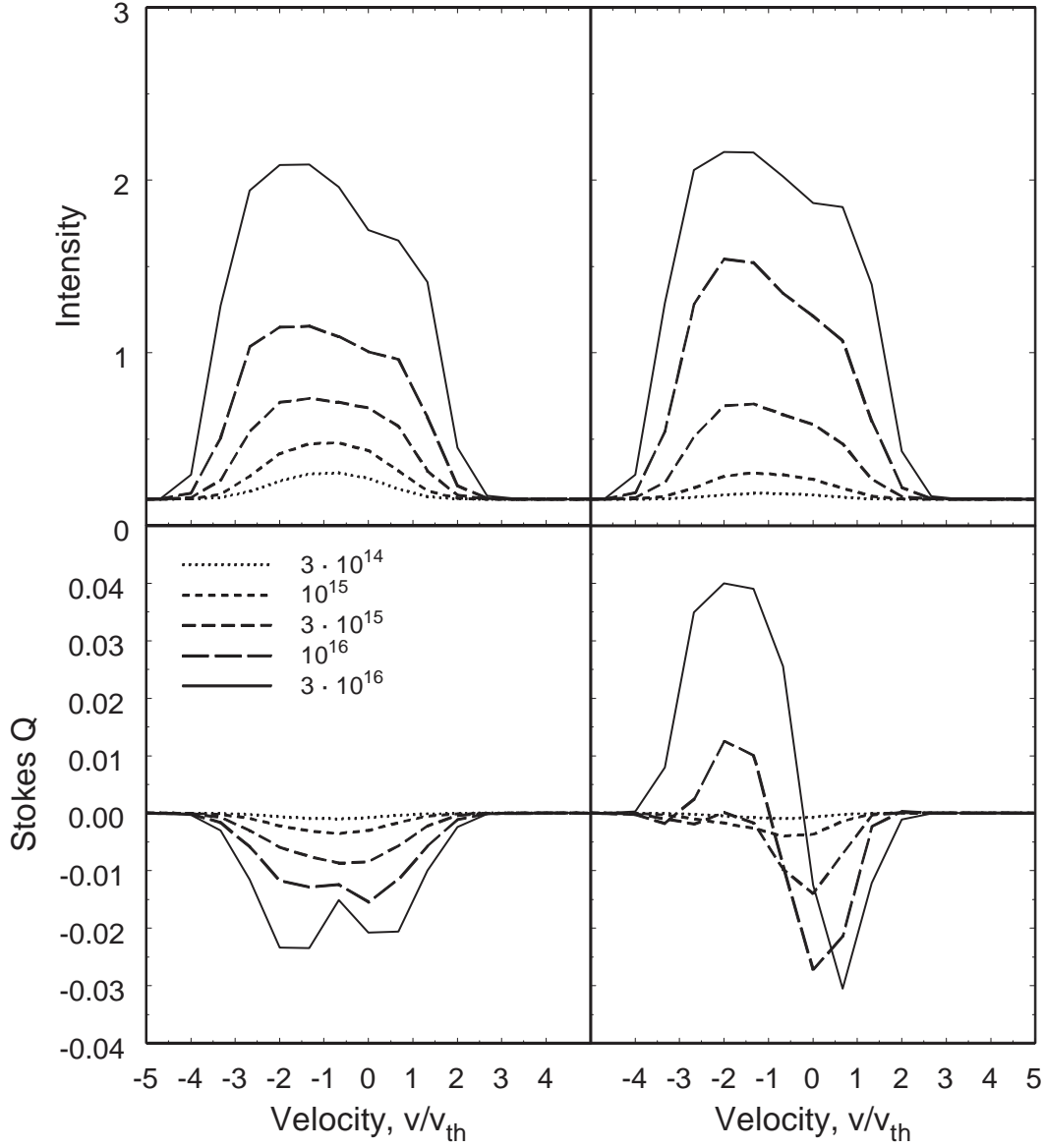}
\caption{Representative intensity and Stokes-Q profiles (arbitrary intensity
units) obtained with the LVG (left panels) and non-local (right panels)
approximations and $v_{\mathrm{A}}/c_{s}=10$ (5th time step).
Profiles are presented for several values of the column density $N_{\rm CO}$ (cm$^-2$) of CO molecules as indicated by the line types.
}
\label{prof015}
\end{figure}

\begin{figure}[tbp]
\includegraphics[scale=.80]{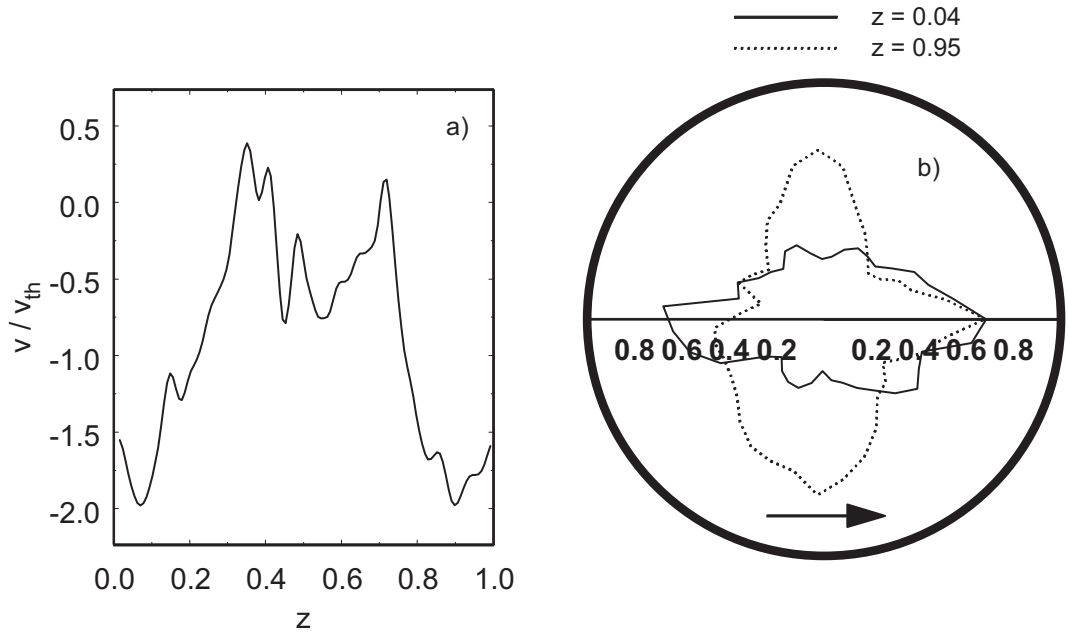}
\caption{(a) The component of the turbulent velocity along the line of sight
and (b) the angular distribution of the effective interaction length $L_{%
\mathrm{eff}}$ (arbitrary units) in the plane of the sky for two locations
along the ray where the contributions to the polarization are orthogonal to
one another. An arrow indicates the direction of the average magnetic field
and $z$ is the distance along the ray measured from the far side of the cube.
}
\label{tauphi}
\end{figure}
\clearpage
The two methods are further compared in Figure~\ref{prof015}, where the line
profiles for a single ray of radiation (the intensity that emerges from a
single grid point on the surface) are given for several values of $N_{%
\mathrm{CO}}$. The specific ray shown is the ray marked with a circle in
Figure~\ref{vecmapnl}. This is one of the cases where the polarization
vector (representing the polarization at the peak of the intensity profile)
is parallel to the average magnetic field in the non-local case and
perpendicular to it in the LVG case. This ray is, thus, not typical but
instead represents a case where the differences between the results with the
two approximate methods are greatest.

The reason for this behavior is indicated in Figure~\ref{tauphi}, where we
show the variations in the LOS component of the turbulent velocity along
this ray. Profiles for Stokes-$I$ and $Q$ of the emergent radiation are
shown in Figure~\ref{prof015}. The spectral line consists of two components
with mutually perpendicular polarizations, as indicated by the sign of $Q$
in bottom left panel (in this Paper, negative $Q$ corresponds to
polarization that is perpendicular to the direction of the average magnetic
field).

Orthogonal polarizations originate in these two velocity components because
of the detailed behavior of the velocity field along the ray. The component
at the line-of-sight (LOS) velocities between about $-0.5v_{\mathrm{th}}$
and $0.5v_{\mathrm{th}}$ which occurs at distances $z$ between $0.3L_{0}$
and $0.7L_{0}$ from the far side of the cube, has polarization perpendicular
to the magnetic field. The component with LOS velocities near $-1.5v_{%
\mathrm{th}}$ originates from gas at two locations along the ray,
specifically, at $z<0.3L_{0}$ and $z>0.7L_{0}$. The region at the far side
of the cube generates emission that is polarized perpendicular to the
magnetic field because the effective interaction length (Eq.~[\ref{leff}])
is largest there along the direction of the average magnetic field (solid
line in Figure~\ref{tauphi}b). This is also the case in the intermediate
part of the ray. Consequently, the optical depth is larger along the
magnetic field. This creates the necessary conditions for overpopulation of
the $m=\pm 1$ substates, resulting in stronger $\sigma $ transitions and
hence polarization that is perpendicular to the magnetic field.

Closer to the near surface of the cube, the structure of the velocity field
changes so that the effective interaction length and the optical depth are
greater in the direction perpendicular to the magnetic field (Figure~\ref
{tauphi}b; dotted line). This causes overpopulation of the $m=0$ substate,
enhanced $\pi $ transitions, and hence a net polarization that is parallel
to the magnetic field. When the emergent radiation consists of two (or more)
velocity components with different polarizations, the resultant polarization
direction is determined by the stronger of the components. As the region
with the ``wrong'' velocity structure occupies a shorter portion of the ray
that does the region with the common velocity structure, it shows up as a
polarization reversal only when the optical depth along the ray is
sufficiently large\textbf{. }That is, there are no reversals if $N_{\mathrm{%
CO}}$ is smaller than some critical value (Stokes-Q is negative for all
profiles with $N_{\mathrm{CO}}\leq 3\times 10^{15}$~cm$^{-2}$ on bottom left
panel of Figure~\ref{prof015}). The number of reversals in a particular cube
increases with increasing $N_{\mathrm{CO}}$ (at the same time, the
percentage polarization decreases). In the case of the ray in Figure~\ref
{tauphi}, the angular distribution of $L_{\mathrm{eff}}$ shown with the
dotted line in Figure~\ref{tauphi}b is only created when the radiative
coupling with remote locations is taken into account. It is not evident from
the angular distribution of the local velocity gradients. Hence, the
polarization reversal does not appear in the LVG approximation (bottom right
panel of Figure~\ref{prof015}).

\subsection{Effect of the Finite Beam Size}

The polarizations shown in Figures~\ref{vecmapnl} and~\ref{vecmaplvg} are
quite high. However, the vectors plotted in these maps represent single rays
and do not reflect the possible effects of averaging over the finite size
for the beam of a realistic telescope. Realistic averaging may cause a
cancellation of the polarization vectors and a decrease in the observed
polarization percentage. Such averaging may also lead to a disappearance of
the reversals in polarization direction that are observed (even for the
strongest magnetic fields) in Figures~\ref{vecmapnl} and~\ref{vecmaplvg}.
Another concern is whether there is some substructure between the rays in
Figures~\ref{vecmapnl} and~\ref{vecmaplvg} that is being missed. We have
thus computed higher resolution maps for selected, representative regions of
the surfaces of the cubes.

\begin{figure}[tbp]
\includegraphics[scale=.80,clip=]{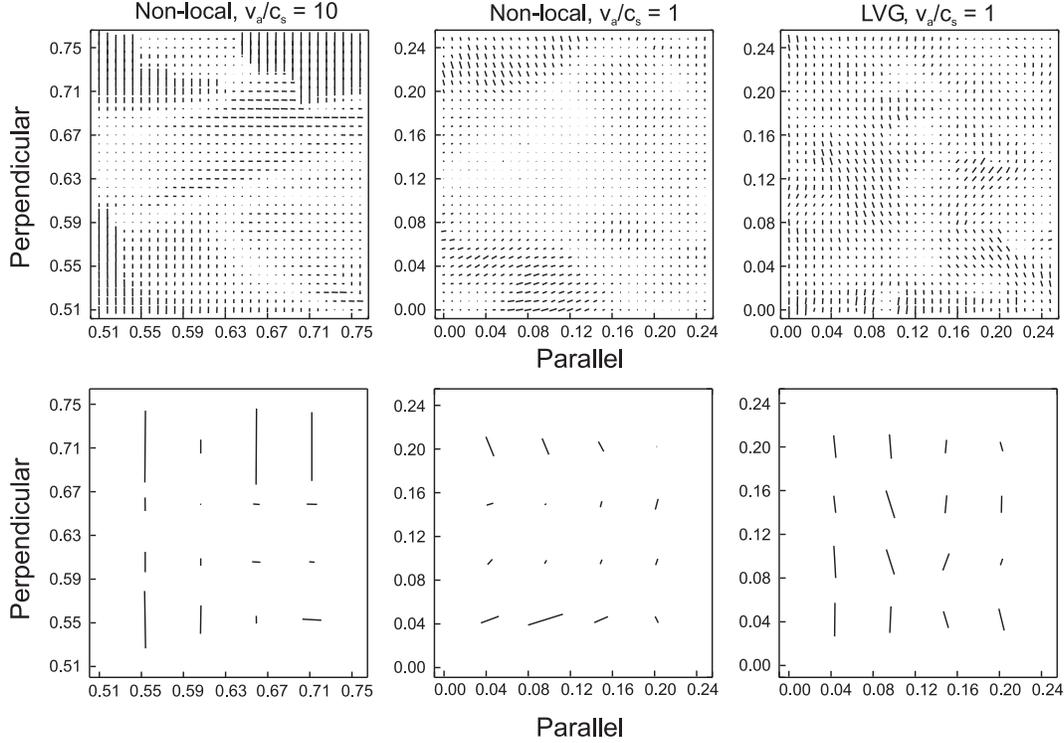}
\caption{High resolution maps showing the polarization vectors for all
emergent rays in $32\times32$ areas of grid points (top row) and with
averaging over $8\times8$ sub-areas (lower row). Left panels---a map for a
region marked with a circle on Figure~\ref{vecmapnl} with $v_{\mathrm{A}%
}/c_{s}=10$. Middle panels---bottom, left corner of the surface of a cube
with $v_{\mathrm{A}}/c_{s}=1$ obtained with the non-local approximation.
Right panels---same region as in middle panels, but for the LVG
approximation. The labels on the axes refer to directions parallel and
perpendicular to the average magnetic field. The longest vector in the top
row of maps corresponds to 2.8\% and, in the lower row of maps, to 1.8\%.}
\label{hires}
\end{figure}

In the upper left panel of Figure~\ref{hires}, we present a high resolution
map for the region centered on the reversed ray that is marked with a circle
in Figure~\ref{vecmapnl}. The map shows the polarization vectors for all $%
32\times 32$ rays emerging from the region of $32\times 32$ grid points. In
lower left panel of Figure~\ref{hires} we show the same map, but averaged
over rays within $8\times 8$ squares to simulate the effect of finite beam
size. Because the initial perturbations include wavelengths that are a
significant fraction of $L_{0}$, the regions exhibiting reversed
polarizations can occur over an appreciable fraction of the surface of the
cube, even in the case of the stronger magnetic field. Their effect on the
observational data will be reduced because of the finite size of the
telescope beam, though the effects will tend to persist until the beam size
reaches approximately $0.1L_{0}$. In the case of the stronger magnetic
field, the regions where these reversals occur are quite localized and beam
averaging will not necessarily reduce the overall polarization when the
telescope beam is centered on most locations on the surface of the MHD cube.
\clearpage
\clearpage
In the case of $v_{\mathrm{A}}/c_{s}=3$ where the fraction of reversed
polarization vectors is about one third, mutually perpendicular
polarizations within one beam are common and the resulting, observed
polarization is decreased significantly by cancellation. This cancellation
is even greater in the case of $v_{\mathrm{A}}/c_{s}=1$. In the center and
right panels of Figure~\ref{hires}, we present high resolution maps for the
lower left corner areas of the maps in Figures~\ref{vecmapnl} and~\ref
{vecmaplvg} for the $v_{\mathrm{A}}/c_{s}=1$ case. The vectors in the center
panels of Figure~\ref{hires} are computed with the non-local approximation,
while the vectors in left panels are computed in the LVG approximation.
Again, the maps in the lower panels are obtained by averaging the rays of
the upper panels within $8\times 8$ squares.

The polarization structure obtained in the LVG calculation is more regular,
and the polarization percentage is higher, than in the non-local
calculation. This difference occurs because the non-local integration
encompasses a significant fraction of the cube, and the probability is high
that integration paths will encounter regions with the ``wrong'' velocity
gradients---which are plentiful in the computations with the weak magnetic
fields. On a significant portion of the surface of these MHD cubes, the
percentage polarization of the emerging rays is negligible. However, there
are still some localized regions with the polarization at a level of 1\%.
The directions of the polarization vectors in these regions reflect the
large-scale structure in the magnetic and velocity fields, rather than the
reversed direction discussed above.

The percentage polarization is decreased by the finite beam size and
averages almost to zero when the beam size is greater than $\sim 0.2L_{0}$.
The polarization is higher in the LVG (than in the non-local) approximation
and tends to be perpendicular to the average magnetic field even in the $v_{%
\mathrm{A}}/c_{s}=1$ case. Hence, the effect of the beam averaging is less
dramatic in the LVG approximation.

In summary, we see that the percentage polarization is not reduced
significantly for modest beam sizes in either the non-local approximation or
in the LVG approximation, and persists at a level of $\sim 1\%$ even for the
weakest magnetic field case that we have considered.

\section{Discussion}

There are now two prevailing views on the evolution of molecular clouds.
According to the so called ``standard'' star formation scenario, molecular
clouds are rather long-lived entities supported against gravity by the
magnetic field \citep{shuaraa}. This support is gradually lost due to ambipolar diffusion,
so that some dense clumps within a parent cloud become gravitationally
unstable and collapse to form protostars. In the alternative turbulent
scenario, molecular clouds are dynamic transient objects---being formed and
destroyed on a timescale which can be only slightly longer than the
dynamical time \citep{mmmlrk}. The existing observational data seem to favor the turbulent
scenario, though the evidence can be ambiguous \citep{mousch2006}.

The polarimetry of molecular lines offers further insights into the nature
of chaotic motions and magnetic fields in star-forming regions as well as in
other astronomical phenomena. The anisotropy of the decaying MHD turbulence
has previously been shown by \cite{ohmas} to reproduce observed polarization
characteristics of OH masers. In this paper, we demonstrate that this
anisotropy can also cause a linear polarization of thermal molecular lines
that is similar in magnitude to what is observed. The highest fractional
polarization is obtained in our calculations when the average magnetic field
is strongest---where the polarization pattern also is most regular. The
number of polarization reversals is greater for magnetic fields of
intermediate strength, and is approximately equal to the number of
polarization vectors in the ``standard'' direction when $v_{\mathrm{A}}/c_{%
\mathrm{s}}=1$.

\begin{figure}[tbp]
\includegraphics[scale=.80,clip=]{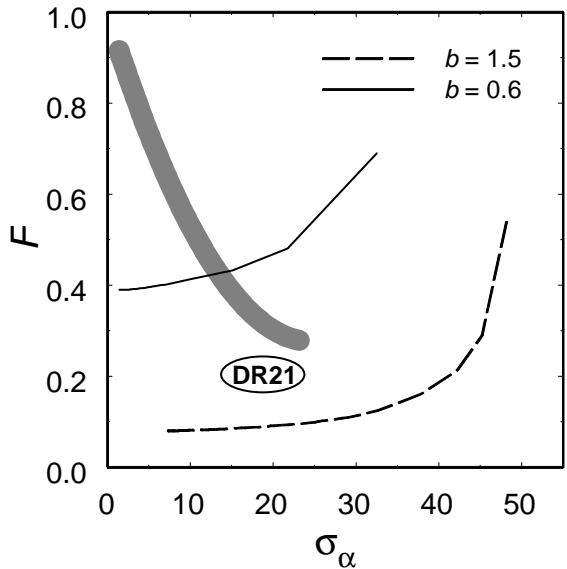}
\caption{Dispersion in the position angles for the polarized emission from dust
vs. polarization reduction factor for models considered in \protect\cite
{dpol} (solid and dashed lines) and in the current paper (thick gray band).
The location of a particular point on the thin lines is determined by the
number of correlation lengths across the cube (increasing from right to
left). The location of a point on the thick band is determined by the value
of $b$ (increasing from top to bottom). The values of $\protect\sigma%
_\protect\protect\alpha$ and $F$ for DR~21(OH) are computed from the
observational data of \protect\cite{lai03}.}
\label{dustpol}
\end{figure}

To see how this result can relate to our previous studies, we consider the
polarization of the far infrared emission by dust that would result from the
same magnetic field distrubutions. In Figure~\ref{dustpol} we reproduce the
left panel of Figure~6 from \cite{dpol}, and plot the dispersion $\sigma
_{\alpha }$ in the position angles of the polarization vectors versus the
polarization reduction factor $F$ (which is a relative measure of the
percentage polarization) using the turbulent MHD fields from the CO
calculations to compute the emission by dust. In our previous studies, we
were mainly interested in the relationship between the regular and irregular
components of the magnetic field. Hence, we used the ratio of the strength
of the random (rms) to the uniform magnetic field 
\begin{equation}
b=B_{\mathrm{rms}}/B_{\mathrm{avg}}
\end{equation}
as a main parameter. In the current study, $b$ is a function of both time
and $v_{\mathrm{A}}/c_{\mathrm{s}}$. We simulated the dust polarization in
the data cubes using the same methods as in \cite{dpol}. The location of a
specific point in Figure~\ref{dustpol} depends on $b$ and on the number of
correlation lengths across the cube. The correlation length does not change
significantly during the time evolution in the current simulations. Thus,
the $\sigma _{\alpha }$ and $F$ computed here depend only on $b$ and lie
along a single curve, denoted with the thick gray band in Figure~\ref
{dustpol}. The bottom part of the band corresponds to lower $v_{\mathrm{A}%
}/c_{\mathrm{s}}$ and/or earlier timesteps. The upper part of the band
corresponds to higher $v_{\mathrm{A}}/c_{\mathrm{s}}$ and/or later timesteps.

It can be seen in Figure~\ref{dustpol} that the turbulent MHD models in our
current study lead to polarization characteristics for emission by dust that
overlap with the results of our computations in \cite{dpol} only for the
smallest values of $v_{\mathrm{A}}/c_{\mathrm{s}}$. On the other hand, the
directions of the polarization vectors of the observed CO emission tend to
be aligned or are smoothly varying---behavior that occurs in our models for
which the regular magnetic fields are stronger (values $b\la0.1$ are typical
for models with $v_{\mathrm{A}}/c_{\mathrm{s}}\geq 3$) than those considered
in our dust polarimetry project.

However, it must be stressed that thermal emission by dust and polarized CO
emission trace regions that are dramatically different. The optical depth of~%
$\sim 1$ that produces the highest CO polarization corresponds to $N_{%
\mathrm{CO}}\approx 10^{15}$~cm$^{-2}$, which is equivalent to molecular
hydrogen column density $N_{\mathrm{H}_{2}}\sim 10^{19}-10^{20}$~cm$^{-2}$
(or $A_{\mathrm{V}}<1)$. This extinction is much smaller than is required to
produce observable dust emission. Thus, even when the dust and the polarized
CO emission are observed in the same object, the dust emission
preferentially probes the inner part of the object while the polarized CO
emission traces the magnetic field in the outer envelope. Our results then
tend to indicate that the magnetic field is more uniform in the outer parts
of such gas clouds than in the more dense core region which is responsible
for the thermal emission by dust.
\clearpage
\clearpage
So far, observations provide only limited opportunities to check our
predictions about the polarization of the CO lines. Unfortunately, some
objects where the polarization of thermal lines has been observed are too
complex to represent a case of ``pure'' MHD turbulence. The most appropriate
is the dataset obtained for DR~21(OH). Data for the dust emission and for the
CO(2--1) transition have been obtained by \cite{lai03}, and for dust
emission and the CO(1--0) transition by \cite{cortes}. The polarization
vectors for the CO(1--0) tend to be parallel to the polarization vectors for
the dust emission, whereas the polarization vectors for the CO(2--1) tend to
be perpendicular to both the dust emission and the CO(1--0) vectors. \cite
{cortes} argued that the CO(2--1) polarization can reflect the influence of
the external radiation and can then be directed parallel to the magnetic
field. At the same time, the CO(1--0) polarization reflects the influence of
the velocity structure and is perpendicular to the magnetic field. This is
exactly the situation that is expected from our current computations. In the
absence of a (strong enough) source of continuum radiation, the polarization
vectors for the dust and the CO should be parallel to one another and
perpendicular to the average magnetic field.

Moreover, while we have no way to relate the magnetic fields in our MHD
cubes to the gravitational stability of the objects that they may represent,
our results are in general agreement with the \cite{cortes} conclusion that
the envelope of DR~21(OH) is highly subcritical. The observation that the
directions of the polarization vectors of the dust and the CO(1--0) are
parallel in this object indicates that they trace the same average magnetic
field and are thus correlated. However, the location of DR~21(OH) in the $\sigma
_{\alpha }-F$ diagram (marked in Figure~\ref{dustpol}) implies the value of $%
b$ is near 1. The presence of a significant irregular component as well as
large scale variations in the magnetic field suggests that the field is not
very dynamically important in the dense part of DR~21(OH), which is traced by the
dust emission. On the other hand, the small position angle dispersion for
the CO(1--0) polarization vectors corresponds to much smaller $b$ values
and, thus, to a regular magnetic field structure that is not affected by gas
motions at either large or small scales.

Another detail in the observations of DR~21(OH) presented by \cite{cortes} that
resembles features in the current computations is the $90^{\circ }$ change
in the direction of the CO(1--0) polarization just to the West of the main
polarization peak at $V_{\mathrm{lsr}}=-10$~km s$^{-1}$ (their Figure~4).
However, since these ``rotated'' vectors are parallel to those of the
CO(2--1) polarization they may just indicate the influence of a compact
continuum source rather than the occurence of the ``reversals'' found in
this study.

Because of the low density of the gas traced by polarized CO emission, it
can be used to probe the nature of the turbulence in molecular clouds more
widely than just in regions of star formation. Such observations are beyond
the reach of current instruments, but may be possible with ALMA---provided
that ALMA has the necessary polarimetric capabilities. When these data
become available, and are combined with the polarimetry data from starlight,
a coherent picture of the large scale velocity and magnetic field structure
in the molecular ISM may emerge.

\acknowledgments
We are grateful to Ya. Pavlyuchenkov for discussions about 
radiative transfer issues, and to C. Gammie and J. McKinney 
for continuing to allow us to use the results of their MHD 
computations. DW acknowledges partial support the RF President
Grant NSh-4820.2006.02. Early work was also supported in part by NSF grant
AST99-88104.

\end{document}